\documentclass[preprint2]{aastex}

\def\ls{\mathrel{\lower0.6ex\hbox{$\buildrel {\textstyle <}
 \over {\scriptstyle \sim}$}}}

\shorttitle{Inverted 20 GHz Emission}
\shortauthors{Boughn \& Pober}

\begin{document}

\title{Evidence for Inverted Spectrum 20 GHz Emission in the 
Galactic Plane}

\author{S. P. Boughn and J. C. Pober}
\affil{Department of Astronomy, Haverford College, Haverford, PA  19041;
sboughn@haverford.edu}

\begin{abstract}
A comparison of a 19 GHz full-sky map with the WMAP satellite K band (23 GHz) map 
indicates that the bulk of the 20 GHz emission within $7^{\circ}$ of the
Galactic plane has an inverted (rising) spectrum with  an average spectral index 
$\alpha = 0.21 \pm 0.05$.  While such a spectrum is 
inconsistent with steep spectrum synchrotron ($\alpha \sim -0.7$) and flat spectrum 
free-free (Bremsstrahlung) ($\alpha \sim -0.1$) emission, it is consistent with various 
models of electric dipole emission from thermally excited spinning dust grains as well 
as models of magnetic dipole emission from ferromagnetic dust grains.  Several 
regions in the plane, e.g., near the Cygnus arm, have spectra with even larger 
$\alpha$.  While low signal to noise of the 19 GHz data precludes a detailed map of 
spectral index, especially off the Galactic plane, it appears that the bulk of the 
emission in the plane is correlated with the morphology of dust.  Regions with higher
23 GHz flux tend to have harder spectra.   Off the plane, 
at Galactic latitudes $7^{\circ} < |b_{II}| < 20^{\circ}$, the spectrum 
steepens to $\alpha = -0.16 \pm 0.15$.  
\end{abstract}


\section{Introduction} \label{intro}

The spatial power spectrum of the cosmic microwave background 
(CMB) constitutes one of the most important observations in cosmology
\citep{ben03a}.  In order to accurately measure the fluctuations in the CMB, it 
is essential to correct for diffuse Galactic emission; this need has led to a keen 
interest in  microwave and far infrared emission of the interstellar medium
(ISM).  Three known souces of emission, synchroton, free-free (bremsstrahlung),
and thermal dust emission (above 60 GHz), have all been identified in observations 
of the CMB \citep{ben03b}.  A decade ago, data from the COBE satellite indicated
that at least some of the microwave continuum below 60 GHz is spatially
correlated with Galactic dust \citep{kog96a,kog96b}.  Both the magnitude and
frequency dependence of this dust-correlated component are inconsistent with 
thermal dust emission while low frequency ($\ls$ 1 GHz) synchroton emission 
is largely uncorrelated with dust. In addition, the lack
of a strong correlation with H$\alpha$ emission led Draine and Lazarian (1998a) 
to argue against the possibilty of the presence of dust-correlated free-free emission 
on energetic grounds.  They instead proposed the mechanism of electric dipole emission 
from thermally excited, spinning dust grains \citep{dra98a,dra98b}, and later, that of 
magentic dipole emission from ferromagnetic grains \citep{dra99}.
Since then, many observations have confirmed the presence of dust correlated, low 
frequency emission \citep{deo97,lei97,deo98,deo99,deo02,muk01,muk02}; although, some 
observations failed to detect this emission \citep{deo00,ham01,muk03}.
The spinning dust spectrum is distinguished by a peak at $\sim$ 20 GHz, a spectral feature 
that has been tentatively indentified in two specific sources \citep{fin02} and in a survey 
of the Galactic plane \citep{fin04b}; although, subsequent observations of one of the former 
two sources, an HII region, have indicated that the emission is consistent with optically 
thin free-free emission with little contribution from spinning dust \citep{dic06}.  
In their analysis of WMAP satellite data, Bennett et al. 
(2003b) suggested a model of ``hard'' synchrotron emission from supernovae
remnants that is correlated with dust but not with ``soft'' synchrotron emission at low 
frequencies ($\ls$ 1 GHz).  While it's likely that ``hard'' synchrotron emission
does contribute to 20 GHz emission in the Galactic plane, the rising spectrum 
reported by Finkbeiner et al. (2004) and in this paper require a substantial contribution
of another source with an even harder (inverted) spectrum.

\section{Method} \label{method}

In this paper, we compare a full-sky 19 GHz map \citep{cot87,bou92} with the three year 
WMAP K and Ka band maps \citep{hin06} in order to determine a spectral index in 
and near the Galactic plane. The WMAP $1^{\circ}$ smoothed K and Ka band intensity maps were 
corrected for CMB fluctuations by subtracting the WMAP Internal Linear Combination (ILC)
CMB map and then convolved with 
the 19 GHz antenna beam pattern, including sidelobes out to 90 degrees, so as to match the 
$3^{\circ}$ angular resolution of the 19 GHz map.  The dipole and CMB were similarly removed 
from the 19 GHz map and all maps were converted to a common 24,576 $1.3^{\circ}  \times 1.3^{\circ}$ 
pixelization in an equatorial quadrilateralized spherical cube projection on the sky \citep{whi92}.  
The mean instrument noise per pixel 
of the 19 GHz data is $\sim$ 2 mK, which is by far the dominant source of statistical noise.  
(The mean instrument noise per pixel in the K and Ka maps is 
$\sim6~\mu$K). The calibration uncertainty of the K and Ka band data is  0.5\% 
\citep{hin06}, while that of the 19 GHz map is 3\% \citep{bou92}; however, the 
relative calibration of the two data sets is known to within 0.6\%, the statistical 
accuracy with which the dipole in the 19 GHz map can be determined.  By matching the
dipoles of the data sets, all data were reduced to the same thermodynamic temperature scale.

Systematic errors will be discussed in detail in \S\ref{sys}.  One potential source of error
is the unknown offset in the 19 GHz map.  An estimate of this offset was determined
by comparing the mean 19 GHz and K band emissions at large Galactic latitudes
($30^{\circ} \le |b_{II}| \le 90^{\circ}$) and requiring the ratio of these
two values to be consistent with a synchrotron spectrum.  This conservative choice is
discussed in \S\ref{sys} along with possible variations of the offset with position in
the sky.  The offset uncertainty in the WMAP maps is on the order of $\sim 4 ~ \mu K$ 
\citep{hin06} and is insignificant in the present analysis.

The ratio of the 19 GHz and K band emission is then determined from a maximum likelihood
(minimum $\chi^2$) fit of this ratio times the K band pixelized data to the 19 GHz data 
in the Galactic latitude region of interest where
\begin{equation}
\chi^2 = \sum_i (y_i - r x_i)^2/{\sigma_i}^2 ,
\end{equation}
$y_i$ is the 19 GHz flux in the $i^{th}$ pixel, $x_i$ is the K band flux, $\sigma_i$
is the 19 GHz noise, and $r$ is the presumed constant ratio of the 19 GHz to 23 GHz
emission.  There is a slight pixel-pixel correlation of instrument noise in the 19 GHz 
data, which we ignore.  Its effect is small and, in any case, can be considered to be 
a systematic effect as discussed in \S\ref{sys}.

\section{Results} \label{results}

Table 1 lists the fit values of the ratio of the 19 GHz to K band (23 GHz) and Ka band
(33 GHz) data and the implied spectral indices in two different Galactic latitude 
bins.  The quoted statisitcal errors are the 1 $\sigma$ quadrature combination
of instrument noise and calibration uncertainty.  The transformation 
of the wideband  K and Ka thermodynamic temperature data to the effective
monochromatic Rayleigh-Jeans temperature necessary to compute the spectral index, $\alpha$,
was effected with
the tabulated parameters in Jarosik et al. (2003).  The 19 GHz data is narrow band, allowing
the use of the standard thermodynamic to Rayleigh-Jeans conversion.  Also listed in Table 1 
are $\chi^2$ and degrees of freedom $\nu$ of the fits.   
\begin{deluxetable}{cccccc}
\tablecaption{Fits for $r$, the thermodynamic temperature ratio, and implied 
spectral index $\alpha$ between 19 GHz and K/Ka bands in two Galactic
latitude bins along with the reduced $\chi^2$ and degrees of freedom, $\nu$,
of the fits. \label{tbl-1}}
\tablewidth{0pt}
\tablehead{
\colhead{Latitude Bin} & \colhead{Bands} & \colhead{r} & \colhead{$\alpha$} & 
\colhead{$\chi^2$/$\nu$} & \colhead{$\nu$}
}
\startdata
$0^{\circ}-7^{\circ}$ &19GHz/K &1.327$\pm$0.010 &0.210$\pm$0.048 &1.244 &2995 \\
$7^{\circ}-20^{\circ}$ &19GHz/K &1.404$\pm$0.032 &-0.156$\pm$0.148 &1.067 &5193 \\
$0^{\circ}-7^{\circ}$ &19GHz/Ka &3.113$\pm$0.023 &-0.160$\pm$0.014 &1.335 &2995 \\
$7^{\circ}-20^{\circ}$ &19GHz/Ka &3.599$\pm$0.082 &-0.434$\pm$0.043 &1.077 &5193 \\
\enddata
\end{deluxetable}

Figure 1 is a plot of 
the 19 GHz and K band thermodynamic temperatures for all 2996 pixels within $7^{\circ}$ of the 
Galactic plane along with the straight line fit, the slope of which is the $r$ listed in
Table 1.  Also shown in Figure 1 are dotted lines corresponding to $\alpha$'s of -0.14 and -0.71, 
the spectral indices of free-free and soft synchrotron emission.
\begin{figure*}
\begin{center}
\includegraphics[scale=1.8]{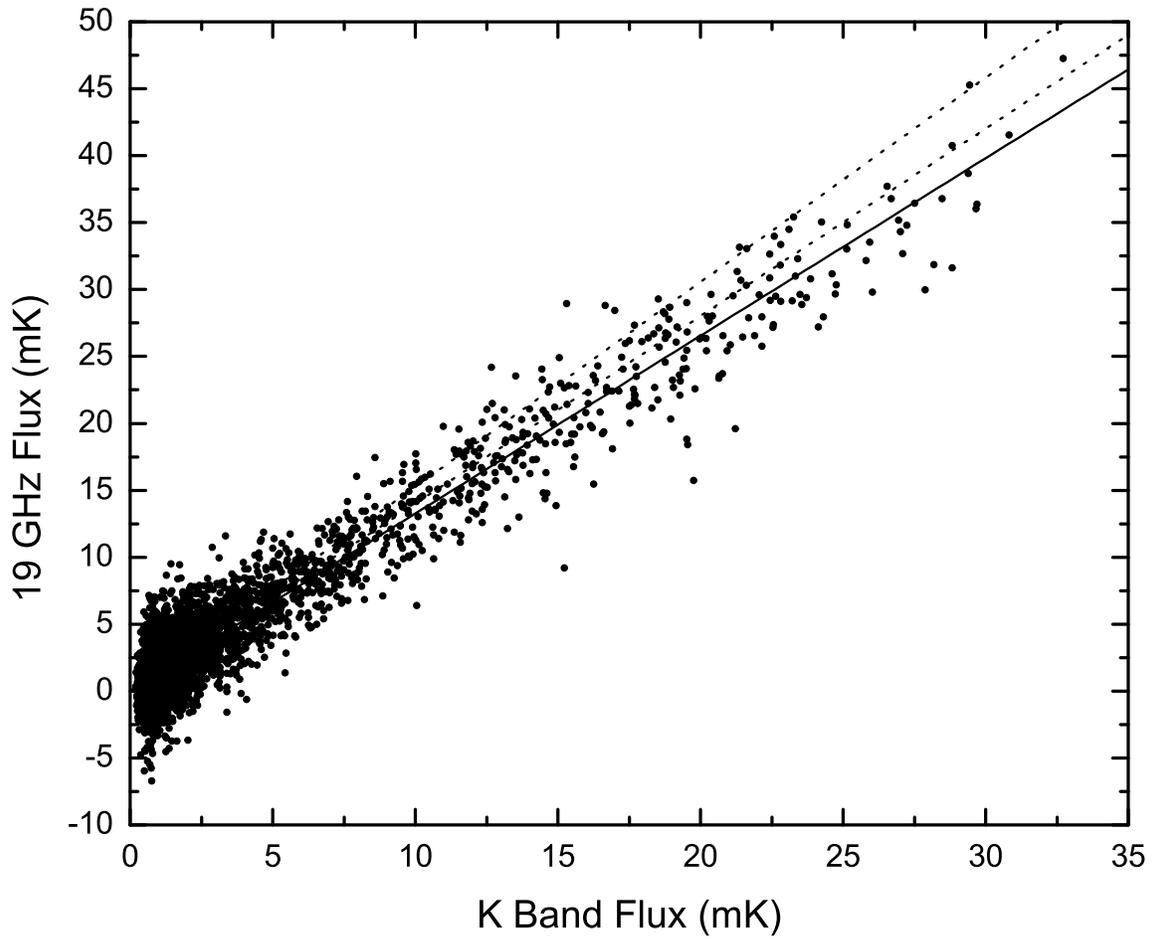}
\caption{The correlation of 19 GHz emission with K band emission for the 2996 pixels
within $7^{\circ}$ of the Galactic plane.  The solid line is the minimum
$\chi^2$ fit to the data while the two dotted lines correspond to spectral indices of 
-0.14 (smaller slope) and -0.71 (larger slope) appropriate for free-free and soft
synchrotron emission.}
\end{center}
\label{fig:cor}
\end{figure*}  
The scatter 
of the points in Figure 1 is dominated by 19 GHz instrument noise and it is clear that most of
the data points have low signal to noise.  This precludes a fine scale map of $\alpha$; however,
averaging pixels on a larger angular scale does allow some measure of the variation of spectral 
index with position.  Figure 2 is a plot of $\alpha$ derived from fits in individual 
$10^{\circ}  \times 10^{\circ}$ regions centered at $5^{\circ}$ intervals along the Galactic plane.
While the errors on many of these points are quite large, there is significant variation 
of the spectral index along the Galactic plane within $\pm 100^{\circ}$ of the 
Galactic center and there are several regions with $\alpha$ significantly larger than the 
value of 0.21 that characterizes the Galactic plane
region as a whole.  Of note is $\alpha \sim 0.77 \pm 0.13$ for the region centered on 
$\ell_{II} \sim 75^{\circ}$ near the Cygnus arm. 
\begin{figure*}
\begin{center}
\includegraphics[scale=1.8]{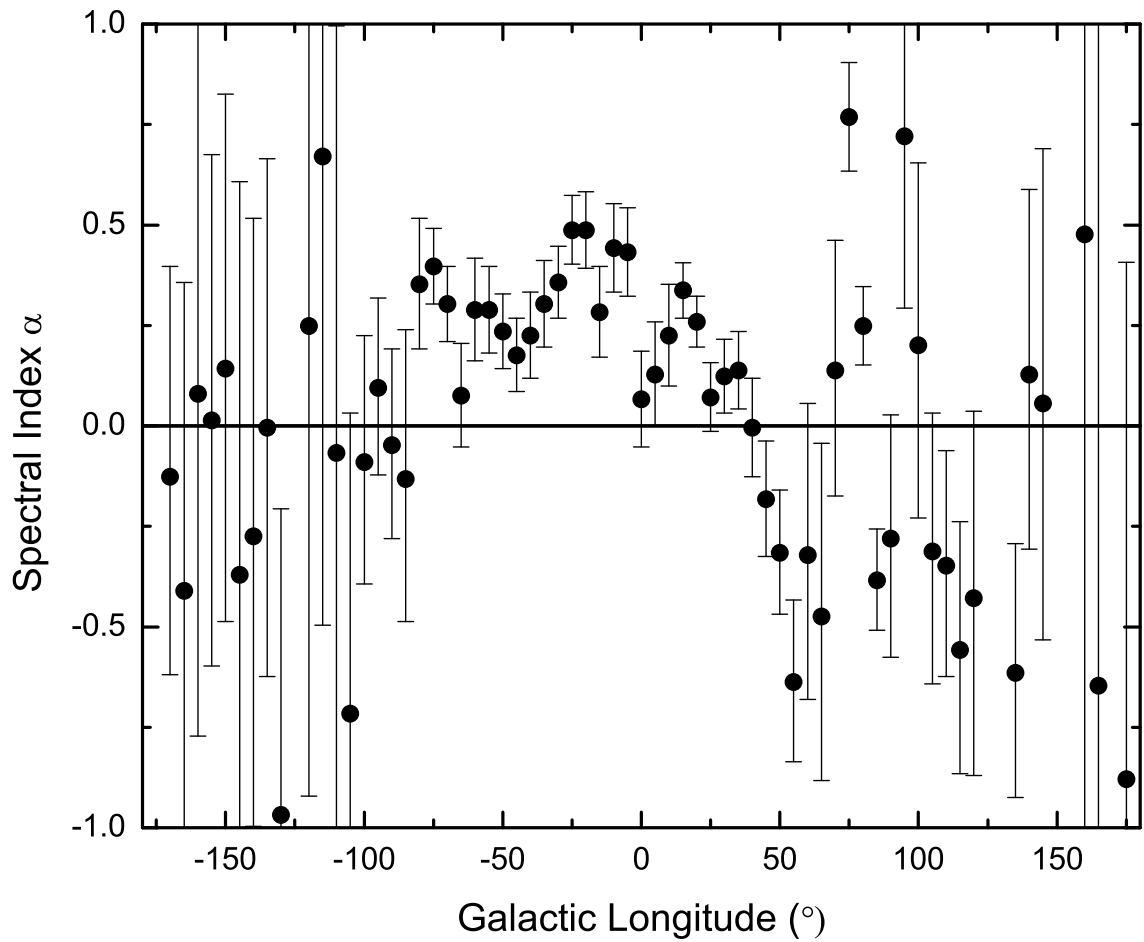}
\caption{The mean spectral index $\alpha$ in $10^{\circ} \times 10^{\circ}$ regions
at $5^{\circ}$ intervals along the Galactic plane.  The error bars represent statistical errors only.}
\end{center}
\label{fig:plane}
\end{figure*}

 As another more qualitative way to illustrate 
the variation of $\alpha$, we smoothed the 19 GHz
and K band maps with an $8^{\circ}$ FWHM Gaussian filter.  Figure 3 is a map of $\alpha$ implied
by the ratio of those pixels in the two maps for which the signal to noise, $S/N = r/\sigma_r$, 
is greater than 10.  The variations of $\alpha$ in Figure 3 are readily idenitified with
corresponding regions in Figure 2.
\begin{figure*}
\begin{center}
\includegraphics[scale=.7,angle=270]{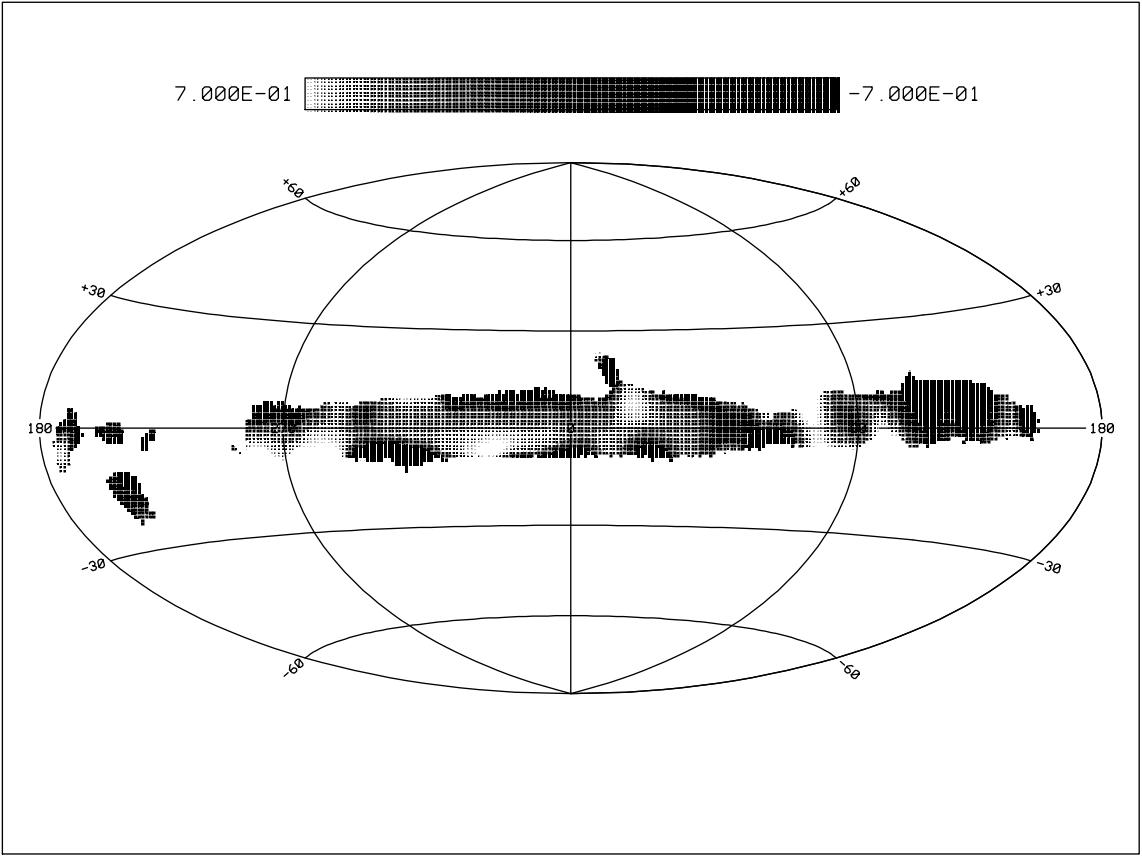}
\caption{Map of the spectral index, $-0.7 < \alpha < 0.7$, deduced from the 
ratio of 19 GHz to K band emission smoothed to a resolution of $8^{\circ}$ FWHM.  
Only those pixels for which the 19 GHz signal to noise exceeds 10 are plotted.}
\end{center}
\label{fig:map}
\end{figure*}

So far we have not addressed the issue of whether the inverted spectrum emission is
associated morphologically with dust in the Galactic plane.  A standard way to do this is to
correlate a dust template with low frequency emission.  An effective 
spectral index for the dust correlated component can 
be inferred from the relative amplitudes of the correlation at two different frequencies.
Following Finkbeiner et al. (2004), we use the WMAP W band (93 GHz) map since 
``it is a good tracer of dust column density.''  The ratio of the linear, minimum $\chi^2$ fits of the 
W band data to the 19 GHz and K band data is then used to deduce an effective spectral index
between 19 and 23 GHz.  The 19 GHz $\sigma_i$'s were used for the K band data to ensure that 
the pixels in the two maps were weighted in the same way.  The values deduced in this way differ 
insignificantly from those of Table 1 while the $\chi^2$s are marginally worse, $\chi^2=1.442$ 
for the $|b_{II}| < 7^{\circ}$ region and $\chi^2=1.093$ for the 
$7^{\circ} < |b_{II}| < 20^{\circ}$ region.
This is suggestive that the emission is, indeed, correlated with dust; however, it is
important to also compare the results of fits to templates of other forms of emission, 
i.e., free-free and synchrotron.  We take the 408 MHz map of Haslam et al. (1981) as a 
synchrotrom template and the $H\alpha$ map compiled by Finkbeiner (2003) and corrected for 
dust extinction by the WMAP team as a free-free template. The fits of these two templates to the 19 GHz 
data result in significantly worse $\chi^2$'s for the two regions; 6.15 and 1.18 
for the synchrotron template fit and 11.85 and 1.30 for the free-free template fit.  
It should be pointed out that
extinction corrections in the Galactic plane are only approximate; however, the corrected
$H\alpha$ map has similar structure to the WMAP MEM-derived free-free map \citep{ben03b}
and so should provide a rough indication of the distribution of free-free emission.  Also,
the 408 MHz map is biased toward steep spectrum, soft synchroton emission and is unlikely to 
trace hard synchrotron emission.  It is well known that the spectral index of 
synchrotron emission is harder ($\sim -0.5$) toward the Galactic plane in regions of active 
star formation \citep{ben03b} and that this emission is correlated with dust.  Therefore, dust
correlated emission includes both spinning dust and hard synchrotron emission.  

For a direct
comparison we performed a three parameter linear fit of the three templates, dust, free-free, and 
synchrotron, to the 19 GHz data.  From the amplitudes of those fits one can estimate the 
relative fraction of the 19 GHz emission that is associated with the three templates as well as
the spectral indices of the three sources of emission (by taking the ratio of the fit amplitudes
with those of a fit to the K band data).  These values are listed in Table 2.  The blank entries
indicate relatively unconstrained $\alpha$ due to low emission.  It should be
noted that the $\chi^2$ of the three parameter fit in the $|b_{II}| < 7^{\circ}$ region 
is only slightly lower than that of the fit to the dust template alone, which is consistent 
with the high estimated
fraction of dust correlated emission in that region.  The $\chi^2$ in the 
$7^{\circ} < |b_{II}| < 20^{\circ}$ region is marginally lower than that of the fit to the
dust template alone and is also consistent with the relatively large fraction of estimated 
synchrotron emission.  For comparison, the $\chi^2$'s of the 19 GHz data in the absence of any
fit are 23.38 and 1.47 for the two latitude bins, an indication that the templates do, indeed, account for most of the
morphology of the emission in the two regions.  Even so, the $\chi^2$'s in Table 2 indicate significant 
structure that is uncorrelated with the three templates.  This is not unexpected since the
19 GHz emission might well have a different dependence on physical parameters than do the templates.
For this reason the relative fractions listed in Table 2 should not be taken too seriously.
However, qualitatively, it seems reasonable to conclude that dust correlated emission
dominates in the plane while dust and soft synchrotron correlated emission are comparable off 
the plane.  The latter is consistent with the off plane $\alpha$ in Table 1. That the implied 
spectral index of the dust correlated emission is larger than that of synchrotron 
correlated emission in this region should be tempered by the large errors in these quantities.
\begin{deluxetable}{cccccc}
\tablecaption{The relative fraction, $f$, of the contribution of different morphological
types to 19 GHz emission and the associated spectral index $\alpha$ 
in two latitude bins. \label{tbl-2}}
\tablewidth{0pt}
\tablehead{
\colhead{Latitude Bin} & \colhead{Morphology} & \colhead{f} & \colhead{$\alpha$} & 
\colhead{$\chi^2$/$\nu$}
}
\startdata
$0^{\circ}-7^{\circ}$ &dust &0.95 &0.20$\pm$0.08 &   \\
   &free-free &0.03 &\nodata & 1.422 \\
   &synch &0.02 &\nodata &    \\
\tableline
$7^{\circ}-20^{\circ}$ &dust &0.50 &-0.05$\pm$0.50 &   \\
   &free-free &0.07 &\nodata & 1.071 \\
   &synch &0.43 &-0.60$\pm$0.64 &    \\
\enddata
\end{deluxetable}

\section{Systematic Errors} \label{sys}

Systematic errors in the analysis fall in three main categories: relative pointing errors, 
differences in beam profiles, and offset errors.  The pointing accuracy of the 19 GHz 
map is $\ls 1^{\circ}$ \citep{cot87}.  That the direction of the dipole in the 19 GHz data differs
from that in the WMAP data by $ 0.7^{\circ} \pm 0.7^{\circ}$ is consistent with this estimate. 
The beam profile of the radiometer used to make the 19 GHz map is well approximated 
in the E-plane out to $90^{\circ}$  and in the H-plane out to by $45^{\circ}$ by the diffraction
pattern of a $TE_{11}$ mode launched from a circular waveguide  \citep{bou90}.   The antenna was tipped 
$45^{\circ}$ from the vertical in the H-plane and radiation shields were employed to attenuate emission
from the earth.  This resulted in an additional attenuation of 20 db beyond $\approx 45^{\circ}$ in 
the forward H-plane, $\approx 90^{\circ}$ in the side E-planes, and $\approx 135^{\circ}$ in the
back H-plane.  The 60 db point occured at $45^{\circ}$ in the H-plane and $90^{\circ}$ in
the E-plane.    The main beam is asymmetrical by about $\pm 13\%$ \citep{bou90}, which results in a 
complicated and not easily modeled variation of angular resolution across the sky, because a given 
point on the sky is observed with more than one antenna orientation.  An approximate beam 
pattern, consisting of an azimuthally averaged diffraction pattern cut off at $90^{\circ}$, 
was convolved with the WMAP maps.  Changing the cutoff to $45^{\circ}$ had a negligible effect 
on the analysis so we conclude that the far sidelobe pattern has little effect on the results.  
The asymmetry of the main beam is relatively small compared to the angular resolution itself and, 
furthermore, the values of $\alpha$ obtained are on even larger angular scales. However, to check 
for sensitivity to both pointing and beam asymmetry errors we
convolved both the 19 GHz and K band maps with an additional $6^{\circ}$ FWHM gaussian. 
The fitted $\alpha$'s for these maps were less by $0.98~\sigma$ and
$0.55~\sigma$ than the values listed in Table 1 for the $|b_{II}| <
7^{\circ}$ and $7^{\circ} < |b_{II}| < 20^{\circ}$ regions respectively. 
We also computed the spectral indices for the $10^{\circ} \times
10^{\circ}$ regions of Figure 2 by taking the ratio of the 19 GHz to K
band fluxes in these regions.  This is equivalent to smoothing with a
$11^{\circ}$ FWHM filter.  The $\alpha$'s deduced this way differed, on
average, from the values in Figure 2 by only $0.52~\sigma$ with regions of relatively low 19 GHz
emission showing an increase in $\alpha$ and those regions with relatively high 19 GHz emission
showing a decrease in $\alpha$.  Both of these small differences in spectral index are consistent 
with a spatial varation of $\alpha$ as discussed below;
however, the general conclusion is that the systematic effects caused by 
errors in the beam profile and pointing do not significantly affect our analysis. 

Another concern is the uncertainty in the value of the unknown offset in the 19 GHz data as
well as the uncertainty in the variation of the offset with position.  It was pointed out above 
that there is a slight correlation in the noises of different pixels which can result in a 
low level systematic pattern in the map.  This can be considered as simply contributing to a 
variable offset in the map.  Our estimated constant offset, $0.15$ mK, is that required 
to make the ratio of 19 to 23 GHz high latitude ($|b_{II}| \ge 30^{\circ}$) emission
consistent with a soft synchrotron spectrum, i.e., $\alpha \sim -0.7$.  If high latitude 
emission has a flatter spectrum, as would be the case for free-free or spinning dust emission,
the implied offset would be larger.  In turn, a larger offset would result in a larger 
$\alpha$ in our fits to the low Galactic latitude
data.  In this sense, the value of the offset we chose is conservative; however, even if the 
effective spectral index at high latitude were as large as in the plane, 
i.e., $\alpha \sim 0.2$, the implied offset would increase by only 0.03 mK and the implied 
$\alpha$ by only $0.014$.

To investigate the large scale variation in the
offset we computed it for separate hemispheres and separate quadrants of the sky, always in 
regions with Galactic latitude greater than $30^{\circ}$.  The offsets deduced from the north, 
south, east, and west hemispheres and the northeast, northwest, southeast, and southwest 
quadrants have an {\it rms} difference from the nominal 0.15 mK of 0.025 mK with a distribution
consistent with instrument noise.  Even if an offset gradient were this large, its effects would
largely cancel out by symmetry in the above analysis of latitude bins.   
Of course, it might be possible 
that the offset near the plane is different than at large latitudes, i.e., a quadrupole effect.  
To check for this we computed the offsets implied by the data in two different latitude bins, 
$30^{\circ} < |b_{II}| < 50^{\circ}$ and  $50^{\circ} < |b_{II}| < 90^{\circ}$.  These two 
offsets differed from our canonical value by 0.01 mK, about half the statistical error in the
their determination.  The statistical error
of the canonical offset is 0.015 mK.  Therefore, we find it unlikely that the
error in the offset of the 19 GHz map is significantly larger than $\sim 0.02$ mK.  
For comparison, the statistical 
errors listed in Table 1 correspond to offset errors of 0.09 mK for $|b_{II}| < 7^{\circ}$ 
bin and 0.03 mK for $7^{\circ} < |b_{II}| < 20^{\circ}$.  We conclude that systematic
error in the offset does not significantly change the results of Table 1.

One might choose to estimate an effective offset simply by including the offset as an 
additional free parameter in the minimum $\chi^2$ fit of Eq. 1.  Doing this for the data 
in the $|b_{II}| < 7^{\circ}$ region yields an offset of $0.38 \pm 0.04$ mK and a spectral
index of $\alpha = 0.33 \pm 0.05$.  The errors are somewhat larger than in Table 1 
because of the correlation of the two parameters.  The difference of this fit offset
with our canonical value is roughly 5 times the statistical error.  The discrepency
is likely due to the fact that $\alpha$ is not constant in the plane.  It is
apparent from Figure 2 that those regions with large K-band and 19 GHz fluxes, i.e.,
those with high signal to noise, tend to have larger $\alpha$ than those with low fluxes.  The
third of the data points in the plane with the largest K-band band fluxes have an average spectral
index of $\langle\alpha\rangle = 0.27 \pm 0.02$, while the third with the lowest fluxes have
$\langle\alpha\rangle = -0.63 \pm 0.15$ (statistical errors only).  The same is true for the 
data of Figure 1.  The third of the points with the highest K-band fluxes have an average 
19 GHz to K-band ratio of $r = 1.32$ corresponding to $\langle\alpha\rangle = 0.24 \pm 0.02$, 
while the third of the points with the lowest K-band fluxes have $r = 1.49$ and 
$\langle\alpha\rangle = -0.54 \pm 0.28$
(statistical errors only).  The result of this variation is that the low K-band flux
points in Figure 1 have elevated 19 GHz fluxes compared to those expected if the slope was 
that determined by the high K-band flux points.  In a linear fit, this would appear as an
additional offset of about the magnitude observed.  We, therefore, conclude that our canonical
offset determined at high Galactic latitudes is the appropriate choice.  A two 
parameter fit to the $7^{\circ} < |b_{II}| < 20^{\circ}$ data yields an offset within 
0.02 mK of the nominal 0.15 mK and an $\alpha$ of -0.06, consistent with that listed in 
Table 1.

There is the possibility that anomalous small angular scale structure in the 19 GHz 
map, of the sort due to correlated noise, could affect the results expressed in 
Figures 2 and 3.  To investigate this effect we subtracted a scaled K band map from the
19 GHz map and smoothed it with an $11^{\circ}$ FWHM Gaussian filter (corresponding to the
area of the bins in Figure 2). A visual inspection of this map reveals the presence of systematic 
patterns even at high Galactic latitudes.  The residual rms fluctuations (after correcting 
for instrument noise) at high latitudes, $|b_{II}| > 30^{\circ}$, is 0.12 mK.  This should
be considered an upper limit on systematic noise at this scale since it is possible that 
there is still some contamination by real Galactic and extragalactic structure as well as
the possibility of a spatially varing spectral index between 19 GHz and 23 GHz.  For comparison, 
the statistical noise of the signals in the $10^{\circ} \times 10^{\circ}$ bins of Figure 2
correspond to effective offset errors in the range 0.17 mK to 0.45 mK with an average of 0.27 mK.
Therefore, the systematic offset errors are smaller than the the errors quoted in Figure 2 and we
conclude that it is unlikely that systematic errors can significantly change these
results.

The $\chi^2$ (see Table 1), of the fit of the K band to the 19 GHz data in the Galactic 
plane, $\chi^2 = 1.244$, while significantly different from unity, 
is not unexpected. The unaccounted for (i.e., not due to instrument noise) 
{\it rms} residuals in the 19 GHz data are equivalent to an average of 0.78 
mK per pixel.   Some portion of the residuals is undoubtedly due to small scale systematic 
structure in the map.  This structure was estimated to be at the level of 0.12 mK when averaged 
over 100 square degrees ($\sim$ 60 pixels) and on the scale of one pixel, it will likely be larger.
Pointing errors might also contribute to the residuals; however, the largest contribution is
most likely due to the variation of $\alpha$ with position as illustrated in Figures 2 and 3, that is, 
constant $\alpha$ is not the best model. If this is the case, then one would expect a larger
such effect in the comparison of the 19 GHz and 33 GHz maps.  That the $\chi^2$ of the
19 GHz to Ka band fit in Table 1 ($\chi^2 = 1.350$) is significantly larger is in agreement with this
expectation.  The $\chi^2$ of the fit in the  $7^{\circ} < |b_{II}| < 20^{\circ}$ 
region is signficantly smaller and corresponds to residual 
structure at the $\sim 0.39~mK$ per pixel level.  
So while the excessive values of $\chi^2$ in Table 1 might temper our conclusions somewhat, 
they are not unexpected and, in any case, the presence of the relatively small residuals
in the fits does not change the conclusions of this paper.

\section{Discussion} \label{discuss}

The frequencies of the two data sets treated in this work are similiar, 19.2 and 22.7 GHz.
The ratio of the the fluxes at the two frequencies, measured in 
thermodynamic temperature, is only 6\% less than the expected ratio for free-free 
emission and 15\% less than the expected ratio for soft synchrotron emission.  However, the
CMB dipole can be measured quite accurately at both frquencies and, as a consequence, the
relative calibrations of the two data sets can be determined to within 0.6\%, which is
sufficient to distinguish between synchrotron, free-free, and spinning dust emission
mechanisms.  It is clear that 20 GHz emission in the Galactic plane is 
inconsistent with a combination of synchrotron and free-free alone and must include a 
significant component of inverted spectrum emission such as that due to spinning dust. 
Consider, for example, emission consisting of spinning dust with a spectral index of 
$\alpha = 1.0$ and hard synchrotron emission with spectral index $\alpha = -0.5$.  
The spectral index of emission consisting of equal amounts of these two components is 
$\alpha = 0.2$, the value determined for the Galactic plane in Table 1.  Therefore, 
our results are consistent with spinning dust accounting
for a significant fraction of the 20 GHz emission in the Galactic plane.
This conclusion is strengthened by our finding that the morphology
of the emission is much more correlated with dust than with the expected morphologies of soft
synchrotron and free-free emission.  From fits to templates of these three sources we 
estimate that the 19 GHz emission within $7^{\circ}$ of the Galactic plane is dominated by
emission correlated with dust.  To be sure, there is significant variation of the spectral index 
as is indicated in Figures 2 and 3, including regions in which the spectral index is 
significantly less than zero.  Finally, we find some evidence that regions of relatively
high K-band flux have harder spectra than those with relatively low flux.  This is also
consistent with the fact that K-band and 19 GHz flux is relatively highly correlated with
dust, which would have a hard spectrum due to the spinning dust mechanism.  

We have only determined the average flux and spectral index of the 
emission near one frquency (20 GHz) and, therefore, have not attempted 
to fit the data to any particular model of spinning dust emission.  Such models have many free
paremeters, e.g., grain size, composition, temperature, electric dipole moment, gas density, 
ionization fraction, etc.
Suffice it to say that some models do have inverted $20$ GHz spectra as we observe and the 
parameters can be adusted to account for the level of emission \citep{dra98a}.  The rising 
spectrum detected from 19 GHz to 23 GHz is in mild conflict with two 
of the four Galactic plane regions investigated by Finkbeiner et al. (2004).  Although 
their analysis also supports the case for spinning dust emission, in these two regions
(those centered at $\ell_{II} = 15^{\circ}$ and $\ell_{II} = 30^{\circ}$)
they detect a falling spectrum from 14 to 23 GHz.  However, the quoted error is
relatively large and a somewhat lower 14GHz flux would be consistent with the data in 
Figure 2.  Their results in the other two regions, $\ell_{II} = 0^{\circ}$ and 
$\ell_{II} = 45^{\circ}$, the former of which has a rising spectrum and the latter a
falling spectrum, agree qualitatively with Figure 2.  Finally, the effective spectral index 
between 19 GHz and 33 GHz, the WMAP Ka band, is $\alpha \sim -0.16$ in the plane and 
$\alpha \sim -0.43$ in the $7^{\circ} < |b_{II}| < 20^{\circ}$ region.  This observation 
is not necessarily inconsistent with spinning dust emission, the spectra of which can turn 
over at frequencies above 20 GHz.

\acknowledgments

We are grateful to Lyman Page who suggested the comparison of the 19 GHz and K band maps
and to Bruce Partridge for helpful conversations. Dave Cottingham constructed the 19 GHz map 
with the help of Ed Cheng and Dale Fixsen.  We acknowledge the use of the Legacy Archive 
for Microwave Background Data Analysis (LAMBDA). Support for LAMBDA is provided by the 
NASA Office of Space Science.

\clearpage

\clearpage

\end{document}